\documentclass[aps,prl,twocolumn,showpacs,superscriptaddress,groupedaddress,notitlepage]{revtex4-1}
\usepackage[english]{babel}
\usepackage{amsmath,amssymb}
\usepackage{graphicx}
\usepackage{dcolumn}
\usepackage{bm}
\usepackage{amsmath}
\usepackage{xcolor}
\usepackage[applemac]{inputenc}
\usepackage{mathtools}

\renewcommand{\vec}{\mathbf}

\newcommand*{\diff}{\mathop{}\!\mathrm{d}}
\renewcommand{\phi}{\varphi}
\newcommand{\dertot}[2]{\frac{\diff #1}{\diff #2}}
\newcommand{\derpar}[2]{\frac{\partial #1}{\partial #2}}

\newcommand{\type}[1]{\uppercase{S}^{#1}}
\newcommand{\vectype}[1]{\mathbf{\uppercase{S}}^{#1}}
\newcommand{\vecconf}[2]{\mathbf{n}^{#1}(#2)}

\begin{document}


\hspace{5.2in} \mbox{Fermilab-Pub-04/xxx-E}

\title[Forecasting transitions in systems with high dimensional stochastic complex dynamics]{Forecasting transitions in systems with high dimensional stochastic complex dynamics:\\A Linear Stability Analysis of the Tangled Nature Model}
\author{Andrea Cairoli}
\email{a.cairoli@qmul.ac.uk}
\affiliation{School of Mathematical Sciences, Queen Mary, University of London, Mile End Road, E1 4NS, UK}
\author{Duccio Piovani}
\email{duccio.piovani@gmail.com}
\author{Henrik Jeldtoft Jensen}
\email{h.jensen@imperial.ac.uk}
\affiliation{Centre for Complexity Science and Department of Mathematics, Imperial College London, South Kensington Campus, SW7 2AZ, UK}
\begin{abstract}
We propose a new procedure to monitor and forecast the onset of transitions in high dimensional complex systems. We  describe our procedure by an application to the Tangled Nature model of evolutionary ecology. The quasi-stable configurations of the full stochastic dynamics are taken as input for a stability analysis by means of the deterministic mean field equations. Numerical analysis of the high dimensional stability matrix allows us to identify unstable directions associated with eigenvalues with positive real  part. The overlap of the instantaneous configuration vector of the full stochastic system with the eigenvectors of the unstable directions of the deterministic mean field approximation  is found to be a good early-warning of the transitions occurring intermittently. 
\end{abstract}


\maketitle
\textit{Introduction} - Many complex high dimensional systems are characterised by intermittent dynamics, where relatively long quiescent periods are interrupted by sudden and quick bursts of activity during which the system suffers hectic rearrangements. These rearrangements can be seen as transitions between metastable states. Examples of abrupt transitions have been identified in a broad range of systems \cite{scheffer2009critical}: in biological ecosystems \cite{scheffer2001catastrophic,scheffer2003catastrophic} transitions from a flourishing to a wild state can occur, in financial markets \cite{may2008complex} endogenous crisis can destabilize an existing balance, in the human brain \cite{litt2002prediction} epileptic seizures signals a switch from a regular to an irregular condition, climate \cite{lenton2008tipping} can exhibit sudden changes both overall or in one of its subsystems, like  when a bloom of harmful algae suddenly forms in the sea \cite{sellner2003harmful}. Due to their widespread occurrence, these transitions have gathered a huge interest in the last decade, with research mainly focused on developing statistical methods to forecast them from the observed time series \cite{scheffer2009early,lade2012early,scheffer2012anticipating} and on the development of a general mathematical framework to describe them \cite{kuehn2011mathematical}. In the present paper we contribute to both efforts by developing a mathematical analysis by use of a paradigmatic model exhibiting intermittent stochastic evolution and by identifying systemic observables that can deliver early-warning of impending transitions. We focus on the Tangled Nature (TaNa) model \cite{tana:article1, tana:article2, tana:article3} of evolutionary ecology. The initial aim of the model was to establish a sound and simple mathematically framework for "punctuated equilibrium", i.e. the observed intermittent mode of macro-evolution.  

The TaNa model is an individual based stochastic model of coevolution. The model's phenomenology is in good agreement with biological observations \cite{LLJ}.  At the microscopic level of individuals the dynamics is unfolding at a smooth constant pace:  agents reproduce, mutate and die at essentially constant rates. On the contrary, at the systemic level the generated ecological network structures jump from one metastable configuration to another (denoted quasi-Evolutionary Stable Strategies or qESS).  We investigate these macroscopic instabilities by performing a Linear Stability Analysis (LSA) of the mean field representation of the dynamics about the actual configurations produced by the full stochastic dynamics.  LSA is obviously a standard procedure to analyze the nature of fixed points for deterministic autonomous equations of motion. Here we develop the method to allow applications to high dimensional {\em stochastic} dynamics. 

We recall that the LSA for a deterministic autonomous system of equations: $\dot{\bf n}={\mathbb F}({\bf n})$ for the time dependent vector ${\bf n}(t)$ consists in first identifying the fixed points ${\bf n}^*$, i.e. the time independent solutions: $\dot{\bf n}^*=0$ of the full non-linear set of equations of motion. One next studies to the first order terms the time dependence of the deviation: $\delta {\bf n}(t) ={\bf n}(t)-{\bf n}^*$  about each of these fixed points: $\delta\dot{\bf n}={\mathbb M}[{\bf n}^*] \delta{\bf n}$. This equation has the solution $\delta\vec{n}(t)=e^{(t-t_0)\mathbb{M}[\vec{n}^{*}]}\,\delta\vec{n}_0$ and the stability of a given fixed point ${\bf n}^*$ is now determined by the properties of the spectrum of the matrix ${\mathbb M}[{\bf n}^*]$, with unstable directions being connected to eigenvalues with positive real parts. 

When we apply this procedure to very high dimensional situations like the TaNa model it is not possible to solve directly the fixed point equation: ${\mathbb F}({\bf n}^*)=0$. Instead, we can use the observed qESS configurations generated by the full stochastic dynamics to approximate ${\bf n}^*$ and perform a LSA of the mean field dynamics about these configurations. To our knowledge this procedure for applying  LSA to high dimensional stochastic dynamics has hardly been attempted before. Only recently, LSA of agent-based models have been studied: in \cite{timpanaro2012connections} the stability properties of the attractors of a generalized Sznajd model are derived from its mean-field formulation, whereas in \cite{olmi2012stability} a similar analysis has been done for a network of pulse-coupled neurons. Neither of these systems, however, exhibits intermittent behavior as we observe in the TaNa.The intermittent dynamics allows us to define a new mean-field based early-warning measure for the occurrence of abrupt transitions.

\begin{figure}[!h]
\centering
\includegraphics[width=\columnwidth,height=2.5cm]{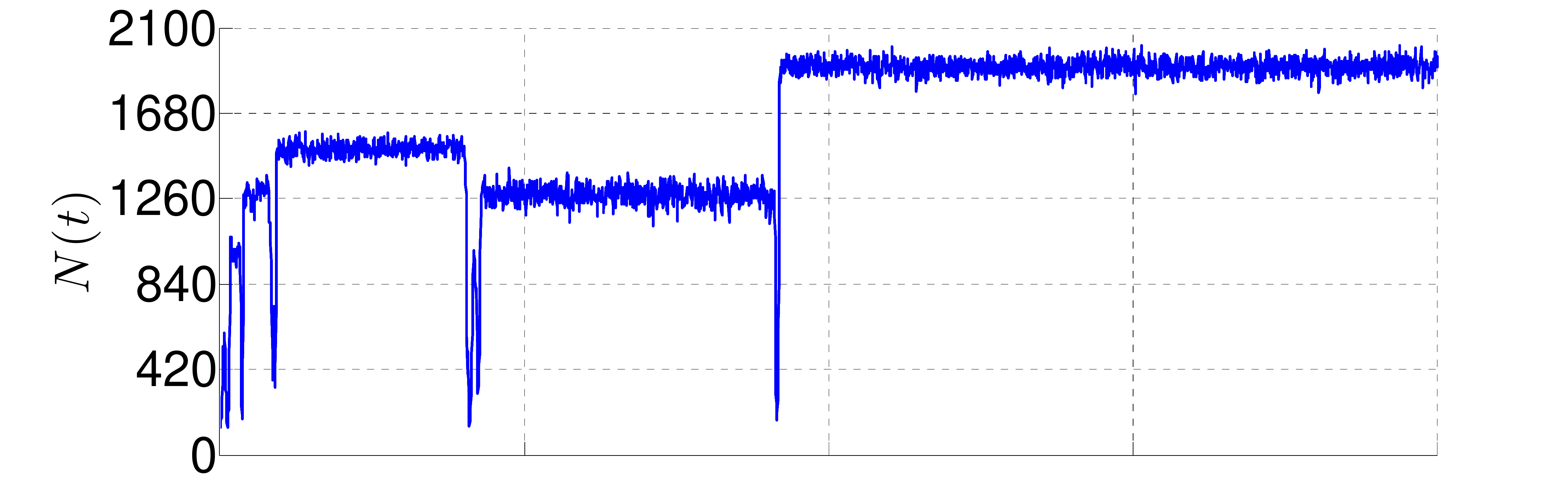}\\
\includegraphics[width= \columnwidth,height=3.5cm]{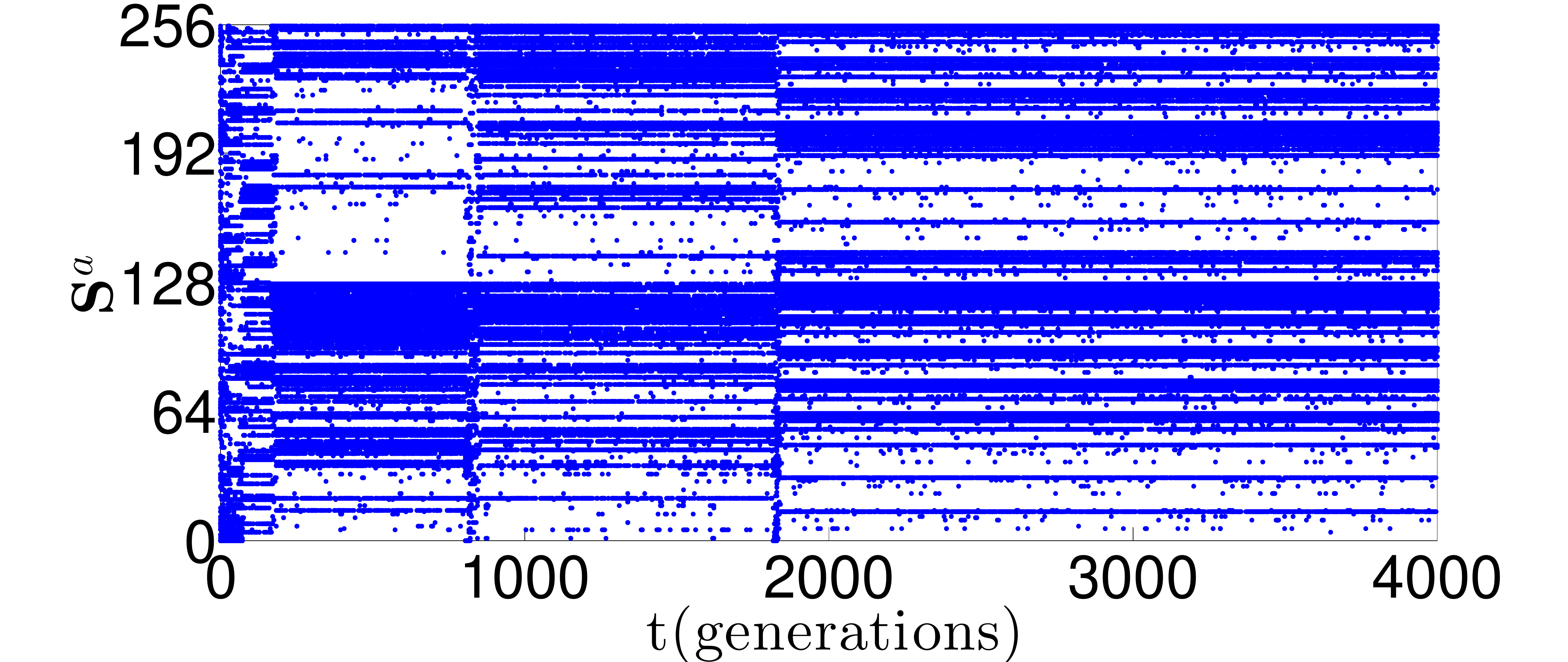}
\caption{Top panel: total population as a function of time (in generations) for a single realization of the TaNa. The punctuated dynamics is clearly visible: quasi-stable periods alternate with periods of hectic transitions, during which $N(t)$ exhibits large amplitude fluctuations. Bottom panel: occupancy distribution of the types. The genotypes are labelled arbitrarily and  a dot indicates a type which is occupied at the time t.}
\label{fig:1}
\end{figure}

\textit{The model} - In the TaNa, an agent is represented by a sequence of binary variables with fixed length L \cite{Higgs/Derrida:article}, denoted as $\vectype{a}=(\type{a}_1,\ldots,\type{a}_L)$, where $\type{a}_{i}=\pm 1$. Thus, there are $2^L$ different sequences, each one represented by a vector in the genotype space: $\mathcal{S}=\{-1,1\}^{L}$. In a simplistic picture, each of these sequences represents a genome uniquely determining the phenotype of all individuals of this type. We denote by $n({\bf S}^a,t)$ the number of individuals of type ${\bf S}^a$ at time $t$ and the total population is $N(t)=\sum_{a=1}^{2^L} n(\vectype{a},t)$. We define the distance between different genomes $\vectype{a}$ and $\vectype{b}$ as the Hamming distance: $d_{ab}=\frac{1}{2L}\sum_{i=1}^{L}|\type{a}_{i}-\type{b}_{i}|$. A time step is defined as a succession of one annihilation and of one reproduction attempt. During the killing attempt, an individual is chosen randomly from the population and killed with a probability $p_{kill}$ constant in time and independent on the type. During the reproduction process, a different randomly chosen individual $\vectype{a}$ successfully reproduces with probability: $p_{off}(\vectype{a},t)=\frac{\exp{(\uppercase{h}(\vectype{a},t))}}{1+\exp{(\uppercase{h}(\vectype{a},t))}}\label{eq:2.3}$, which depends on the occupancy distribution of all the types at time $t$ via the weight function:
 \begin{equation}
\uppercase{h}(\vectype{a},t)=\frac{k}{\,\uppercase{n}(t)}\sum_{\vectype{b}\in\,\mathcal{S}}\mathbf{J}(\vectype{a},\vectype{b})n(\vectype{b},t)-\mu\uppercase{n}(t).
\label{eq:1}
\end{equation}   
In Eq. (\ref{eq:1}), the first term couples the agent $\vectype{a}$ to one of type $\vectype{b}$ by introducing the interaction strength $\mathbf{J}(\vectype{a},\vectype{b})$, whose values are randomly distributed in the interval $\left[-1,+1\right]$. For simplification and to emphasize interactions we here assume: $\mathbf{J}(\vectype{a},\vectype{a})=0$. The parameter $k$ scales  the interactions strength and $\mu$ can be thought of as the carrying capacity of the environment. An increase (decrease) in $\mu$ corresponds to harsher (more favourable) external conditions. The reproduction is asexual: the reproducing agent is removed from the population and substituted by two copies $\vectype{a}_1$ and $\vectype{ a}_2$, which are subject to mutations. A single mutation changes the sign of one of the genes: $\type{\gamma}_{i}\rightarrow -\type{\gamma}_{i}$ with probability $p_{mut}$. Similarly to a Monte Carlo sweep in statistical mechanics, the unit of time of our simulations is a  \emph{generation} consisting of $N(t)/p_{kill}$ time steps, i.e. the average time needed to kill all the individuals at time $t$. These microscopic rules generate intermittent macro dynamics. The system is persistently switching between two different modes: the qESS states and the transitions separating them. The qESS states are characterized by small amplitude fluctuations of N(t) and stable patterns of occupancies of the types (Fig. \ref{fig:1}, respectively top and bottom panel). However, these states are not perfectly stable and configurational fluctuations may trigger an abrupt transition to a different qESS state. The transitions consist of adaptive random walks in the configuration space while searching for a new metastable configuration and are related to  high amplitude fluctuations of N(t). 

\textit{Linear stability analysis} - We now describe the mean field deterministic approximation of the TaNa. The macroscopic  configuration is given by $\vec{n}(t)=(n_{1}(t),\ldots,n_{{2^L}}(t))\in (\mathbb{N}\cup\{0\})^{2^{L}}$ and evolves according to the continuous time mean field equation
\begin{equation}
\label{eq:4}
\dertot{\vecconf{}{t}}{t}=\frac{1}{\uppercase{n}(t)}\mathbb{T}[\vecconf{}{t}]\vecconf{}{t},
\end{equation} 
where 
\begin{align}
\{\mathbb{T}&[\vecconf{}{t}]\}_{ab}=\,(\,p_{off}^{b}(\vec{n}(t))(2p_0-1)-p_{kill})\,\delta_{ab} \notag\\
&+2\,p_{off}^{b}(\vec{n}(t))(p_{mut})^{\uppercase{l}d_{ab}}(1-p_{mut})^{\uppercase{L}(1-d_{ab})}(1-\delta_{ab}). \label{eq:3.3}
\end{align}
Although being an approximation of the real stochastic dynamics \cite{durrett1994importance}, we find that Eqs. \eqref{eq:4} is   able to capture the low frequency variation of the discrete time model. 

As mentioned above we cannot directly derive the nontrivial solutions of the fixed points' equation: $\mathbb{T}[\vec{n}^{*}]\vec{n}^{*}=0$, so instead we use the approximated configuration $\bar{\vec{n}}_{stoc}\simeq \vec{n}^*$,  which is obtained from the full stochastic dynamics as a time average during a qESS. Due to the fluctuations occurring during a qESS (See Fig. \ref{fig:1}) $\bar{\vec{n}}_{stoc}$ is only approximately stationary:  $T[\bar{\vec{n}}_{stoc}]\bar{\vec{n}}_{stoc}\simeq0$. The linearized equation for the deviations away from  $\bar{\vec{n}}_{stoc}$ becomes: $\derpar{}{t}\delta\vec{n}=\mathbb{M}[\bar{\vec{n}}_{stoc}]\,\delta\vec{n}$ and the stability of a qESS is given by the spectrum of eigenvalues of $\mathbb{M}[\bar{\vec{n}}_{stoc}]$. How effective an unstable eigendirection is in destabilising the configuration ${\bf n}(t)$ will depend on the overlap between the deviation from $\bar{\vec{n}}_{stoc}$: $\delta \vec{n}(t)={\bf n}(t)-\bar{\vec{n}}_{stoc}$ and the unstable directions. During the qESS we therefore introduce the following instability indicator:      
\begin{equation}
\label{eq:6}
Q(t)=\max_{\lambda \in Sp^{+}(\mathbb{M}[\bar{\vec{n}}_{stoc}])} \left|e^{\lambda}\left\langle (\vec{n}(t)-\bar{\vec{n}}_{stoc}),\vec{e}_{\lambda}\right\rangle\right|
\end{equation} 
where the eigenvalues $\lambda$ and the correspondent eigenvectors $\vec{e}_{\lambda}$ of $\mathbb{M}[\bar{\vec{n}}_{stoc}]$ can be computed numerically for high dimensions (in our case with the Intel DGEEV routine). $Sp^{+}(\mathbb{M}[\bar{\vec{n}}_{stoc}])$ refers to the eigenvalues with positive real part and the brackets denotes the scalar product. For the TaNa we verified numerically that the stable and the unstable sub-spaces are orthogonal.  $Q(t)$ simply measures the maximal expected growth of $\delta\vec{n}(t)$ during the time interval $\Delta t=1$. 

\textit{Procedure and Results} - We monitor the system in real time. To understand when a qESS is established, we average the occupation vector $\vec{n}(t)$ over time windows of $\Delta T = 100$ time units (i.e. generations in the case of the TaNa) to obtain $\bar{\vec{n}}_{stoc}$. We then check if the system is stationary, i.e. $T[\bar{\vec{n}}_{stoc}]\bar{\vec{n}}_{stoc} \simeq 0$, repeating the process until the condition is satisfied. Once a qESS has been reached  we linearize about the configuration $\bar{\vec{n}}_{stoc}$ and compute $Q(t)$ and the instantaneous deviation: $\| \delta \vec{n} (t) \|= \| \vec{n}(t) -\bar{\vec{n}}_{stoc} \|$. Below we demonstrate that the indicator $Q(t)$ is able to monitor and even forecast the onset of a transition out of the current qESS. The transition shows up directly as an unbounded sudden growth of $\| \delta \vec{n} (t) \|$. Once a transition out of the current qESS has occurred, we average again $\vec{n}(t)$ to establish the new quasi stable configuration $\bar{\vec{n}}_{stoc}$.

\begin{figure}[!htb]
\includegraphics[width=\columnwidth,height=5.2cm]{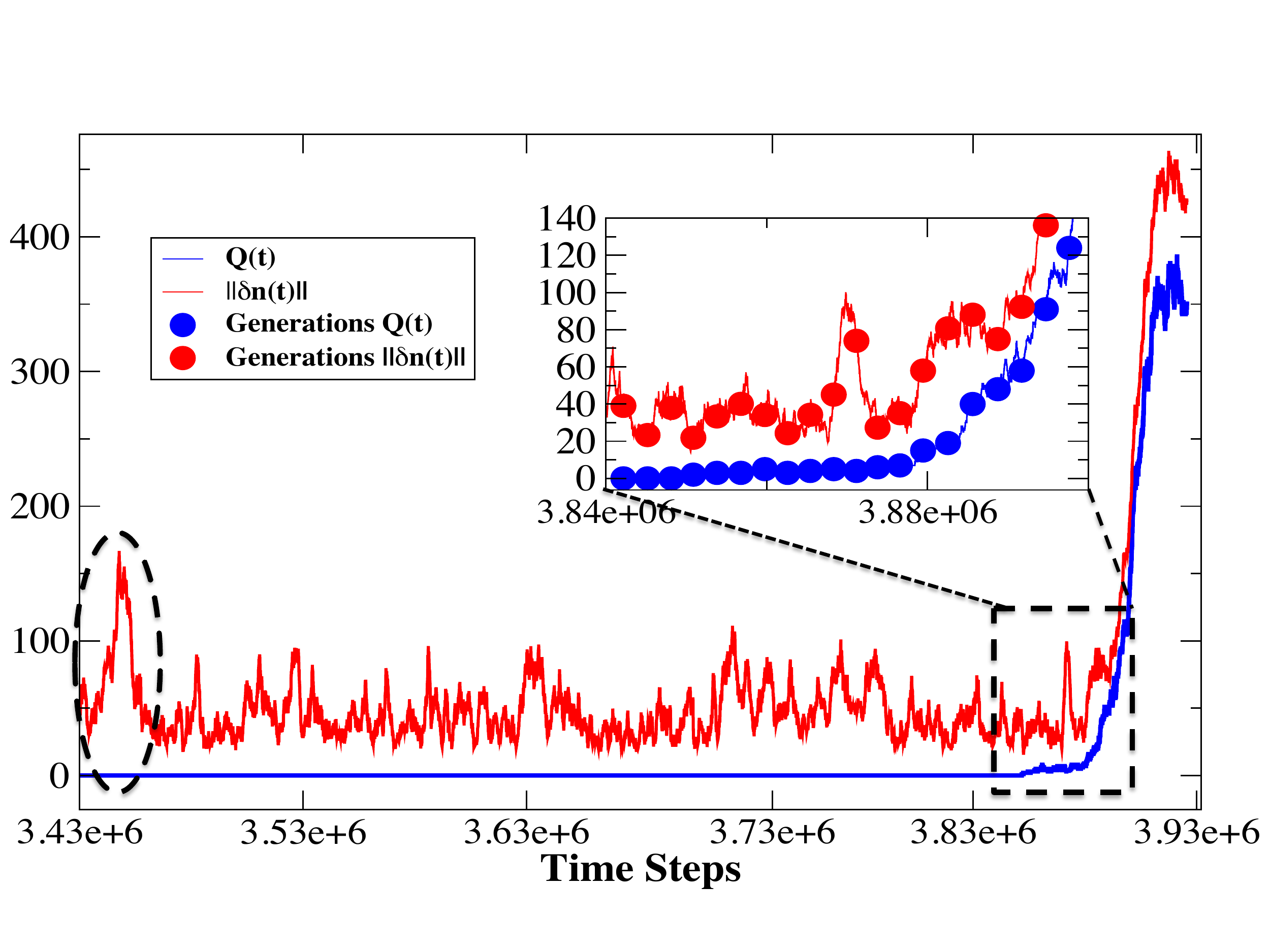}
\caption{Typical behavior of $Q(t)$ and $\| \delta \vec{n} (t) \|$ in a single run of the TaNa in time steps. Clearly $Q(t)\simeq0$ even for more rare strong fluctuations (dashed circle) inside the qESSs, whereas it starts to increase rapidly before the actual transition. In the inset, we zoom on the transition and indicate with markers the points observed at the coarse-grained level of generations. Notice that between two generations many time steps (events) are present.}
\label{fig:2}
\end{figure}

In Fig.\ref{fig:2} we show $Q$ as a function of the microscopic time steps (blue curve).
We observe that $\| \delta \vec{n}(t) \|$ fluctuates during the qESS. In contrast $Q$ only grows when a transition is about to occur. Typically $Q$ starts to increase several generations prior to the transition corresponding, in this particular case, to thousands of single update events.   

\begin{figure*}[!htb]
\includegraphics[width=17cm,height=8.cm]{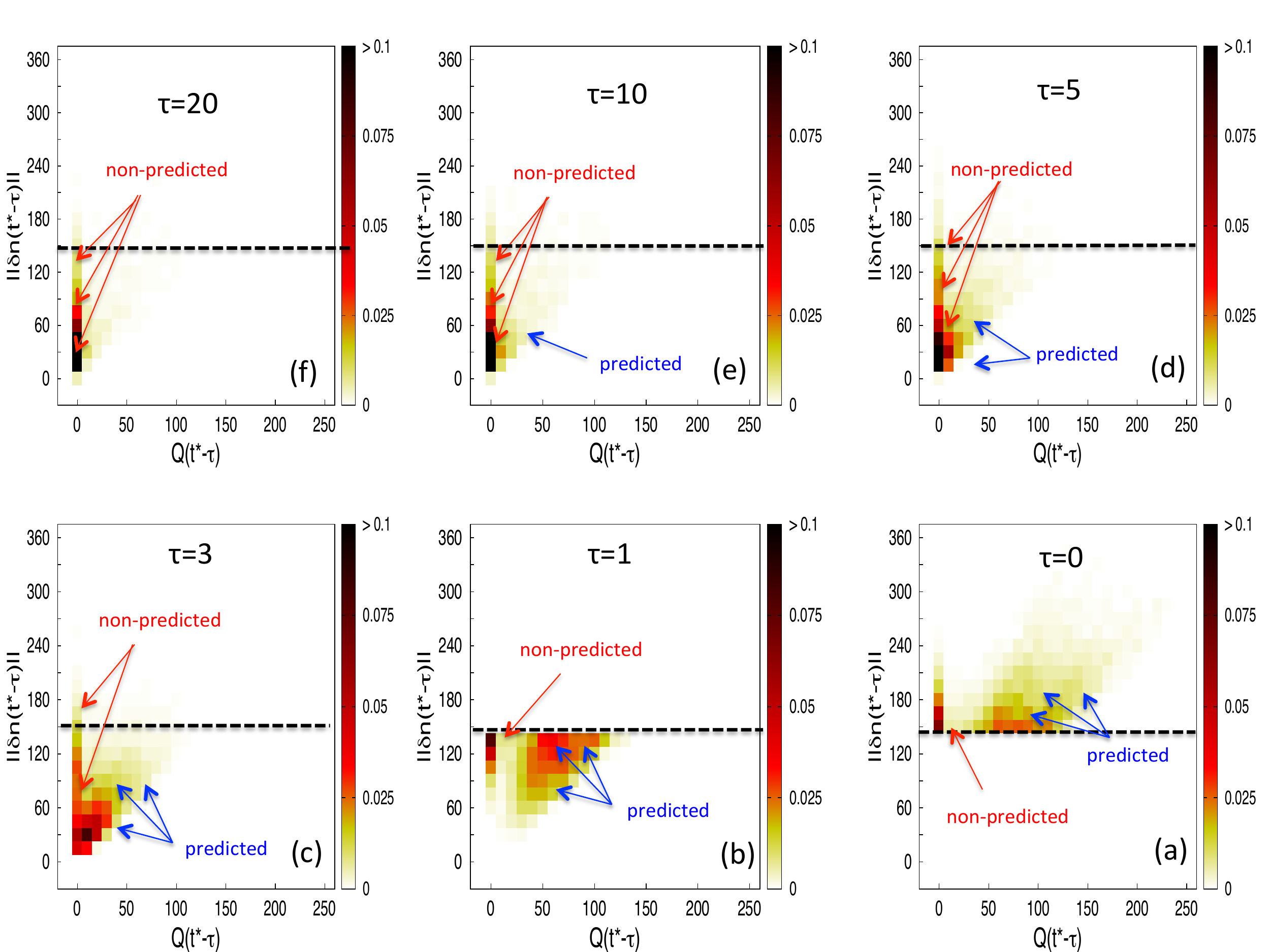}
\caption{2D distribution $P(\| \delta\vec{n}(t^*-\tau)\|,Q(t^*-\tau))$ averaged over 13000 transitions for different values of $\tau$. The predictive power of Q is evident: typical fluctuations inside the qESSs are not signaled by Q (panels (e-f)), whereas dangerous perturbations leading to a transition are recognized by the increasing of Q away from zero (panels (a-d)). This is already seen for $\tau=5$, which is still remarkably far from the transition. Examples of predicted/non predicted transitions are then shown with arrows in panels (d-a). The other plots can be interpreted in a similar way.}\label{fig:3}  
\end{figure*}

To understand the relation between  $\| \delta {\bf n} \|$ and $Q$ we show in Fig. \ref{fig:3} the joint probability density $P(\| \delta\vec{n}(t^*-\tau)\|,Q(t^*-\tau))$ for $\tau$ generations before the time $t^*$ of the transition. We identify $t^*$ as the time when the condition: $\| \delta\vec{n}(t^*) \|>d$ holds persistently for at least the next 10 consecutive generations for a fixed threshold $d=150$ corresponding to the typical upper bound for the amplitude of the fluctuations inside the qESSs. From the way the region of largest support move in the $Q-\|\delta {\bf n}\|$ plane as the transition is approached we see to what extent monitoring $Q$ allows one to predict the transition. Note that a significant support for values of $Q$ larger than about 10 starts to develop from  around $\tau=5$. At these times the deviation $\| \delta\vec{n}\|$ is still most often below the inherent qESS fluctuation level of 150. We may encounter situations where Q gives a false signal, by increasing significantly in correspondence to small amplitude perturbations of $\vec{n}(t)$. Remarkably, we can see that these events happen with low probability, thus not affecting significantly the performance of the Q measure. 

\begin{figure}[!htb]
\includegraphics[scale=0.21]{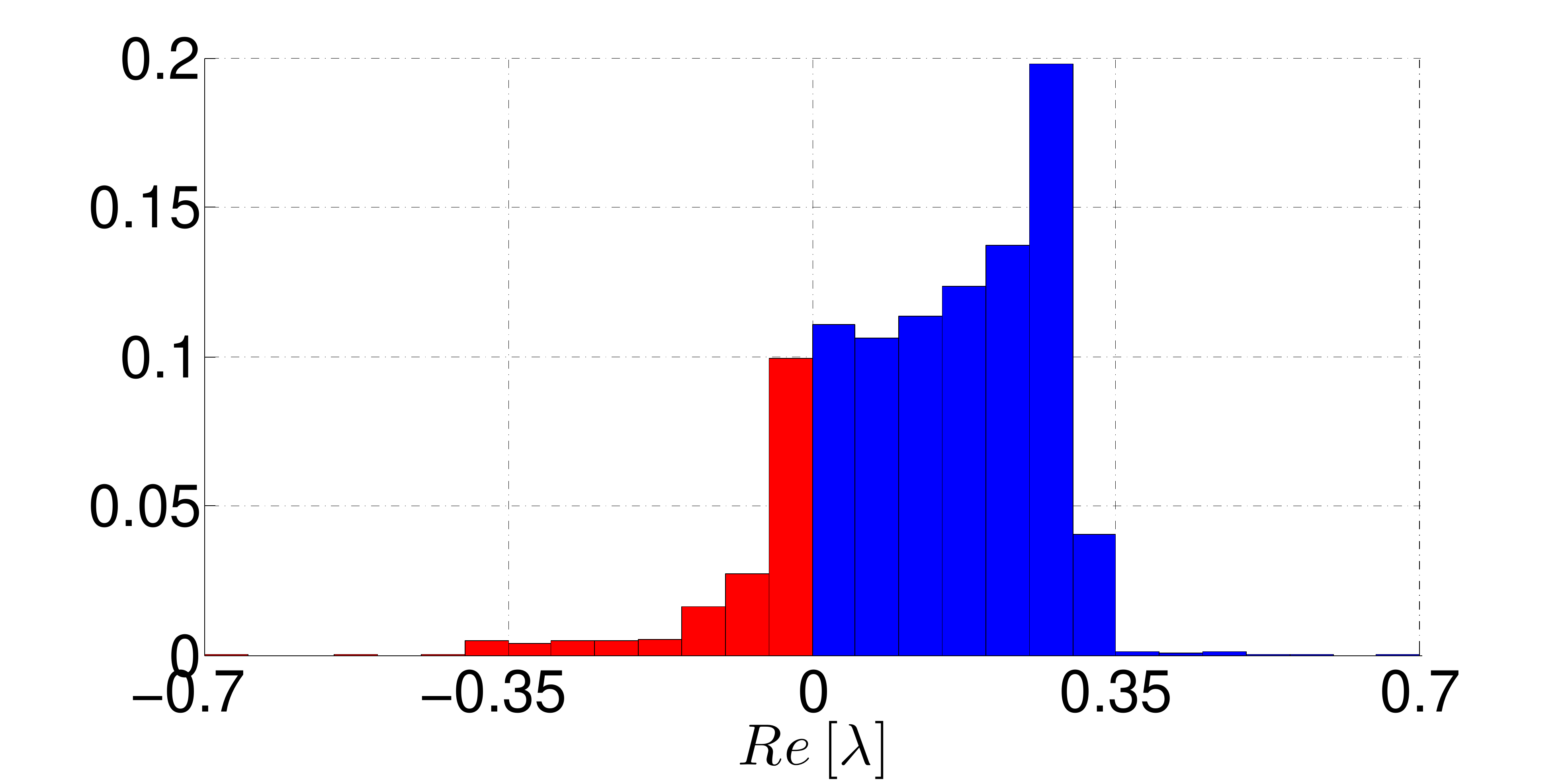}
\caption{Distribution of the real part (red/blue boxes for negative/positive one) of the eigenvalues correspondent to eigendirections with maximum overlap with $\delta \vec{n}(t)$ at the beginning of a transition. The distribution is clearly dominated by the unstable eigenspace, but a significant probability $(\approx 17\%)$ of weak stable eigenvalues is found.}\label{fig:4}
\end{figure}

Finally, our success rate in predicting transitions  is approximately 85-87$\%$, however non-predicted transitions do occur and are related  to a non-vanishing probability that a direction which is weakly stable (negative eigenvalues close to zero) of the mean field can trigger a transition. This is shown in Fig. \ref{fig:4}, where the distribution of the real parts of the eigenvalues responsible for the transitions is plotted. This purely stochastic phenomenon explains why we find with non vanishing probability transitions together with $Q\simeq0$ (see Fig. \ref{fig:3}, panel (a)).  

\textit{Conclusions} - 
We combined deterministic mean field analysis with stochastically generated configurations to  develop a measure capable of forecasting abrupt transitions in the TaNa model. We believe that the procedure outlined here can be applied to other high dimensional complex systems, including for example economy or neuronal systems, when sufficient data sampling is possible to establish the effective interaction matrix (the ${\bf J}$ matrix above) of the fluctuating dynamics. Its mean field approximation can then allow the monitoring of $Q$. With the rapid development towards big-data sampling capacity in many areas of science this scenario becomes increasingly a possibility.


\textit{Acknowledgments} - This work was supported by  the European project CONGAS (Grant FP7-ICT-2011-8-317672)

\textit{Contributions} - 
A.C. and D.P. contributed equally to this work. 
\vspace*{-1.cm}

\end{document}